\documentstyle[preprint,tighten,version2,aps]{revtex}
\begin{document}
\draft
\preprint{IFUP-TH 56/94}
\begin{title}
Large-N phase transition in lattice 2-d principal chiral models.
\end{title}
\author{Massimo Campostrini, Paolo Rossi, and Ettore Vicari}
\begin{instit}
Dipartimento di Fisica dell'Universit\`a and I.N.F.N.,
I-56126 Pisa, Italy
\end{instit}
\begin{abstract}
We investigate the large-$N$ critical behavior of 2-d lattice chiral models
by Monte Carlo simulations of $U(N)$ and $SU(N)$ groups
at large $N$. Numerical results confirm strong coupling analyses,
i.e. the existence of a large-$N$ second order phase transition
at a finite $\beta_c$.
\end{abstract}

\narrowtext

\section{Introduction}
\label{intro}

Strong coupling studies of lattice 2-d principal
chiral models, with the standard nearest-neighbour interaction
\begin{equation}
S_L = -2 N \beta \sum_{x,\mu}
{\mathop{\rm Re}}{\mathop{\rm Tr}}\left[ U(x)\,U^\dagger(x{+}\mu)\right]\;,
\;\;\;\;\;\;\;\;\;\;\;\;\;\beta\;=\;{1\over NT}\;,
\label{action}
\end{equation}
have shown evidence of a large-$N$ phase transition
at a finite $\beta_c$, separating the strong coupling and
the weak coupling regions~\cite{Green,chiralPRL}.
An analysis of the
$18^{\rm th}$ order $N=\infty$ strong coupling series of the specific heat
showed a second order critical behavior
\begin{equation}
C \sim  |\beta - \beta_c |^{-\alpha} \,,
\label{Ccrit}
\end{equation}
with the following estimates of $\beta_c$ and $\alpha$:
$\beta_c = 0.3058(3)$ and $\alpha = 0.23(3)$~\cite{chiralPRL,chiralSC}.
This critical phenomenon  is somehow
effectively decoupled from the continuum limit ($\beta\rightarrow \infty$),
indeed dimensionless ratios of physical quantities
are reproduced with great accuracy
even for $\beta < \beta_c$~\cite{chiral,chiralPRL}.

A critical behavior at $N=\infty$ is also present in 1-d lattice chiral models,
where
at $N=\infty$ the free energy is piecewise analytical with a third order
transition between the strong coupling and weak coupling
domains~\cite{Gross}. In these models the parameter $N$ plays a role analogous
to the volume in ordinary systems with a finite number of degrees of freedom
per site,
and the double scaling limit describing the simultaneous approach
$N\rightarrow \infty$ and $\beta\rightarrow\beta_c$ is shown to be equivalent
to finite size scaling of 2-d spin systems close to the
criticality~\cite{Brezin,Heller}.

In this paper we investigate the above large-$N$
critical phenomenon by Monte Carlo simulations,
that is by extrapolating, possibly in a controlled manner, numerical
results at sufficiently large $N$, in the same spirit of the double
scaling limit technique developed in the studies of 1-d matrix models.
We performed Monte Carlo simulations of $SU(N)$ and $U(N)$ models for several
large values of $N$, studying the approach to $N=\infty$.
Some $SU(N)$ Monte Carlo results at large $N$
were already presented in Ref.~\cite{chiral}.
Since $SU(N)$ and $U(N)$ models are expected to have the same large-N limit,
$U(N)$ Monte Carlo results provide further information and check
of the $N\rightarrow\infty$ behavior of lattice principal chiral models.

In the continuum limit $SU(N)$ and $U(N)$ 2-d lattice actions
should describe the same theory even at finite $N$,
in that the additional $U(1)$ degrees of freedom
of the $U(N)$ models decouple. The $U(N)$ lattice action,
when restricting ourselves to its $SU(N)$ degrees of freedom,
represents a different regularization of the $SU(N)\times SU(N)$
chiral field theory.
One loop calculations in perturbation theory give
the following $\Lambda$-parameter ratios
\begin{equation}
{\Lambda_{\overline {MS}}\over \Lambda_L^{^U}}\;=\;
\sqrt{32}\,\exp\left( {\pi\over 2}\right)\;,
\end{equation}
\begin{equation}
{\Lambda_L^{^{SU}}\over \Lambda_L^{^U}}\;=\; \exp\left( {\pi\over
N^2}\right)\;,
\end{equation}
where $\Lambda_L^{^U}$ and $\Lambda_L^{^{SU}}$ are respectively the
$\Lambda$-parameters
of the $U(N)$ and $SU(N)$ lattice actions~(\ref{action}).

The fundamental group invariant correlation function of $SU(N)$ models is
\begin{equation}
G(x) \;=\; {1\over N} \langle\,{\rm Tr} \,[ U^\dagger (x) U(0) ]\,
\rangle \;.
\label{GSUN}
\end{equation}
Introducing its lattice momentum transform $\widetilde{G}(p)$,
we define the magnetic susceptibility $\chi=\widetilde{G}(0)$,
and the second moment correlation length
\begin{equation}
\xi_G^2 = {1\over4\sin^2\pi/L} \,
\left[{\widetilde G(0,0)\over\widetilde G(0,1)} - 1\right]\;\;\;.
\label{xiG}
\end{equation}

In the $U(N)$ case we consider two Green's functions.
One describes the propagation of $SU(N)$ degrees of freedom:
\begin{eqnarray}
G(x) &=& {1\over N} \langle\,{\rm Tr} \,[ {\hat{U}}^\dagger (x) {\hat{U}}(0)
]\,
\rangle\;,\nonumber \\
{\hat{U}}(x)&\equiv& {U(x)\over ({\rm det} \,U(x))^{1/N}}\;.
\label{GUN}
\end{eqnarray}
The other describes
the propagation of the $U(1)$ degrees of freedom associated with
the determinant of $U(x)$:
\begin{equation}
G_{\rm d}(x) \;=\; \langle\,\left( {\rm det} \,[ U^\dagger (x) U(0)
]\right)^{1/N}\,
\rangle\;,
\label{GUND}
\end{equation}
{}From the Green's functions $G(x)$ and $G_{\rm d}(x)$ we can define the
corresponding
magnetic susceptibilities $\chi$, $\chi_{\rm d}$ and second moment correlation
lengths $\xi_G$, $\xi_{{\rm d}}$.

At finite $N$, while $SU(N)$ lattice models do not have any singularity at
finite $\beta$,
$U(N)$ lattice models should undergo a phase transition,
driven by the $U(1)$ degrees of freedom corresponding to the
determinant of $U(x)$,
and following a pattern similar to the 2-d XY model~\cite{Green2}.
The mass propagating in the determinant channel $M_{\rm d}$ should vanish
at a finite value $\beta_{\rm d}$ and stay zero for larger $\beta$.
Then for $\beta > \beta_{\rm d}$
this sector of the theory decouples from the other ($SU(N)$) degrees
of freedom, which are those determining the continuum limit
of principal chiral models for $\beta\rightarrow\infty$.
We recall that
the 2-d XY model critical behavior is characterized by a sharp approach to the
critical point $\beta_{XY}$ ( the correlation length grows exponentially),
a line of fixed point for $\beta > \beta_{XY}$, and a finite specific heat
having
a peak for a $\beta < \beta_{XY}$ (see e.g. Ref.~\cite{Gupta}).

\section{Numerical results.}
\label{NR}
\subsection{The Monte Carlo algorithm.}

In our simulations we used local algorithms containing overrelaxation
precedures. In the $SU(N)$ case, we employed the Cabibbo-Marinari
algorithm~\cite{Cabibbo}
to upgrade $SU(N)$ matrices by updating their $SU(2)$ subgroups,
chosen randomly among the ${N(N-1)\over 2}$ subgroups acting on each $2\times
2$
submatrix.
At each site the $SU(2)$ subgroup identified
by the indices $i,j$ ($1\le i < j \le N$)
was updated with a probability $P={2 \over N-1}p$, so that
the average number of $SU(2)$ updatings
per $SU(N)$ site variable was $\bar{n}=p N$.
In our simulations we always chose $p \lesssim 1$,
decreasing $p$ when increasing $N$.
We used $p\simeq 1$ at $N=9$, $p\simeq 2/3$ at $N=15$ and
$p\simeq 1/2$ at $N=21,30$.
The extension to the $U(N)$ case is easily achieved
by updating, beside $SU(2)$ subgroups, $U(1)$ subgroups.
In our simulations
we upgraded the $U(1)$ subgroups identified by the diagonal elements of the
$U(N)$ matrix.
The $SU(2)$ and $U(1)$ updatings were performed by a mixture
of over-heat-bath algorithm~\cite{PV} (90\%) and standard
heat-bath (10\%).
At fixed parameter $p$, the number of operations per site increases
as $N^2$ at large $N$.

The above algorithm experiences a critical slowing down in $N$, that is
keeping the correlation length fixed
the autocorrelation time grows with increasing $N$. This effect is partially
compensated
by a reduction of the fluctuations of group invariant quantities
when $N$  grows. In the $U(N)$ simulations the quantities related to the
determinant
channel are subjected to large fluctuations, causing large errors in the
measurements.

In Tables~\ref{UNtable} and \ref{SUNtable} we present Monte Carlo data
respectively
for the $U(N)$ and $SU(N)$ simulations.
Finite size systematic errors
in evaluating infinite volume quantities should be smaller than the statistical
errors
of all numerical results presented in this paper.

\subsection{Numerical evidence of a large-N phase transition.}

Lattice chiral models have a peak in the specific heat
\begin{equation}
C\;=\; {1\over N} { {\rm d} E\over {\rm d} T}
\end{equation}
which becomes sharper and sharper with increasing $N$.
In Figs.~\ref{CUN} and ~\ref{CSUN} we plot the specific heat
respectively for the $U(N)$ and $SU(N)$ models.
Such a  behavior of the specific heat
should be an indication of a phase transition for $N=\infty$
at a finite $\beta_c$.
The positions of the peaks $\beta_{peak}$
in $SU(N)$ and $U(N)$  converge from opposite directions,
restricting the possible values of $\beta_c$ to
$0.304 \lesssim \beta_c \lesssim 0.309$.
Notice that Monte Carlo data for $\beta\lesssim \beta_c\simeq 0.306$
approach, for growing $N$, the resummed
$18^{\rm th}$ order large-$N$ strong coupling series of the specific
heat~\cite{chiralSC};
in this region, as expected by strong coupling considerations, the convergence
of $U(N)$ models is faster.

A more accurate  estimate of the critical coupling $\beta_c$
can be obtained by using a finite $N$ scaling Ansatz
\begin{equation}
\beta_{peak}(N) \;\simeq\; \beta_c\,+\, cN^{-\epsilon}\;,
\label{FNS}
\end{equation}
in order to extrapolate $\beta_{peak}(N)$ to $N\rightarrow \infty$.
The above Ansatz is suggested by the idea that the parameter
$N$ may play a role quite analogous to the volume in the ordinary
systems close to the criticality.
This idea was already exploited in the study of 1-d matrix
models~\cite{Carlson,Brezin,Heller}, where double scaling limit turned out
to be very similar to finite size scaling in a two-dimensional critical
phenomenon.
Substituting $L\rightarrow N$ and $1/\nu \rightarrow \epsilon$,
Eq.~(\ref{FNS}) becomes the well-known finite size scaling relationship derived
in the context of the renormalization group theory.
Furthermore the exponent $\epsilon$ should be the same in the $U(N)$ and
$SU(N)$ models,
in that it should be a critical exponent associated to the $N=\infty$ phase
transition.
Notice that the function $\beta_{peak}(N)$ in Eq.~(\ref{FNS}) is considered at
infinite
space volume.

In the study of ordinary critical phenomena the
reweighting technique~\cite{Ferrenberg},
turns out to be very efficient to determine quantities like the position of the
specific heat peak.
In our work we could use this technique only for $N=9$,
since for larger $N$  the reweighting
range around the point where the simulation is performed
turned out to be much smaller than the typical $\beta$ interval
of our simulations. For $N\geq 15$ $\beta_{peak}(N)$ data and their
errors were estimated from the specific heat data reported in the
Tables~\ref{UNtable} and \ref{SUNtable},
supported by the direct measurements of the specific heat derivatives at each
$\beta$.

Our estimates of $\beta_{peak}$ at $N=9,15,21$ for $U(N)$ and $N=9,15,21,30$
for $SU(N)$
fit very well formula (\ref{FNS}).
By a fit with four free parameters,
$\beta_c$, $\epsilon$, $c_{_{U(N)}}$ and $c_{_{SU(N)}}$, we found
\begin{eqnarray}
&&\beta_c \;=\; 0.3057(3)\;,\;\;\;\;\;\;\;\;\;\;\;\;\;\;\;\;\;
\nonumber \\
&&\epsilon \;=\; 1.45(8)\;.
\label{resfit}
\end{eqnarray}
In Fig.~\ref{bpeak} the fit result is compared with the $\beta_{peak}(N)$ data.
A fit with two independent $\epsilon$ exponents, $\epsilon_{_{U(N)}}$ and
$\epsilon_{_{SU(N)}}$ gave compatible results, but larger errors.
Notice that this Monte Carlo  estimate of $\beta_c$ is in agreement with
the determination (\ref{Ccrit}) coming from strong coupling computations.

We checked the finite $N$ scaling Ansatz (\ref{FNS})
in the similar context of the large-$N$ phase transition of 1-d lattice $U(N)$
chiral models with free boundary conditions, where the critical point $\beta_c$
and
the critical exponents $\nu$ and $\alpha$ are known: $\beta_c=1/2$, $\nu=3/2$
and $\alpha=-1$.
We computed the position of the specific heat peak at finite
$N$ finding the asymptotic behavior (\ref{FNS}) with
$\epsilon=2/3$. Details on these calculations are given in the Appendix.
As already mentioned, from standard finite size scaling arguments
the critical exponent $\epsilon$ should be related to the critical exponent
$\nu$:
$\epsilon=1/\nu$.
Notice that the critical exponents $\nu$ and $\alpha$
satisfy a two-dimensional hyperscaling relation: $2\nu=2-\alpha$.
In 1-d lattice chiral models the number $d_e=2$ of effective dimensions
of the large-$N$ critical phenomenon is related to the fact that
the double limit $N\rightarrow\infty$ and $\beta\rightarrow\beta_c$
is equivalent to the continuous limit of a two-dimensional
gravity model with central charge $c=-2$.

Since the large-$N$ phase transition of the 2-d lattice chiral models
is of the second order type, its behavior  cannot be found in the
classification of  double
scaling limits of Refs.~\cite{Kazakov,Gross}, which  are parametrized by a
central charge $c<1$
implying $\alpha < 0$.
Moreover, unlike 1-d lattice chiral models,
the interpretation of the  large-$N$ phase transition of 2-d lattice chiral
models
as an effective  $d_e=2$ ordinary critical phenomenon does not seem to be
valid:
in fact,  if $\epsilon=1/\nu$, by substituting our estimates of $\alpha$ and
$\epsilon$
in the hyperscaling relation $d_e=(2-\alpha)\epsilon$ we would obtain
$d_e=2.6(2)$.
A more general thermodynamic inequality would give
 $d_e \geq (2-\alpha)\epsilon$~\cite{Stanley}.

Monte Carlo data of $\chi$ and $\xi_G$ for $\beta\lesssim \beta_c$
compare very well with
the large-$N$ strong coupling series of $\chi$ (up to $15^{\rm th}$ order)
and $\xi_G$ (up to $14^{\rm th}$ order)~\cite{chiralPRL}.
Fig.~\ref{xi}, where $\xi_G$ is plotted versus $\beta$,
shows that data approach, with growing $N$,
the curve obtained by resumming the strong coupling series of
$\xi_G$~\cite{chiralSC},
and in particular the $U(N)$ data, whose convergence is faster, are in
quantitative
agreement.

Large-$N$ numerical results seem to indicate that all physical quantities,
such as $\chi$ and $\xi_G$, are well behaved functions of the internal energy
$E$ even
at $N=\infty$~\cite{chiral}.
Therefore as a consequence of the specific heat divergence at $\beta_c$,
the $N=\infty$ $\beta$-function $\beta_L(T)\equiv -a {\rm d}T/{\rm d}a$
should have a non-analytical zero at $\beta_c$,
that is $\beta_L(\beta)\sim |\beta-\beta_c|^\alpha$ in the neighbourhood of
$\beta_c$.
By defining a new temperature proportional to the energy~\cite{Parisi},
this singularity disappears, and one can find good agreement
between the measured mass scale and the
asymptotic scaling predictions in the ``energy'' scheme
even for $\beta < \beta_c$, where strong coupling expansion is expected
to converge~\cite{chiral}.
In fact strong coupling computations show
asymptotic scaling with a
surprising accuracy of few per cent~\cite{chiralPRL}.

In the $U(N)$ case, a Kosterlitz-Thouless
phase transition driven by the determinant is expected
at $\beta_{\rm d} > \beta_{peak}$ for each finite $N$.
Our data seem to support this picture, indeed after the peak of $C$,
the magnetic susceptibility $\chi_{\rm d}$  and the second moment correlation
length
$\xi_{{\rm d}}$ defined from the determinant correlation function~(\ref{GUND})
begin to grow very fast.
In Fig.~\ref{chid} we plot $\chi_{\rm d}$ versus $\beta$.
Green and Samuel argued
(using strong coupling and weak coupling arguments) that the large-$N$ phase
transition
is nothing but the large-$N$ limit of the determinant phase transition present
in the $U(N)$ lattice models~\cite{Green2,Green3}.
According to this conjecture, in the large-$N$ limit
$\beta_{\rm d}$ and $\beta_{peak}$ should both converge to $\beta_c$,
and  the order of the determinant phase transition would change
from an infinite order of the
Kosterlitz-Thouless mechanism to a second order with divergent specific heat.
The large-$N$ phase
transition of the $SU(N)$ models could then be explained by  the fact that
the large-$N$ limit of the $SU(N)$ theory should be the same
as the large-$N$ limit of the $U(N)$ theory.
Our numerical results give only a partial confirm of this scenario,
we can just hint from the behavior of $\chi_{\rm d}$ and $\xi_{{\rm d}}$ with
growing $N$
that the expected phase transition is moving toward $\beta_c$.
The large-$N$ strong coupling series of the mass $M_{\rm d}$
propagating in the determinant channel has been calculated up to $6^{\rm th}$
order,
indicating a critical point, determined by the zero of the $M_{\rm d}$ series,
slightly
larger than our determination of $\beta_c$: $\beta_{\rm d}(N=\infty)\simeq
0.324$~\cite{Green2}.
This discrepancy could be explained either by the shortness of the strong
coupling series
of $M_{\rm d}$, or by the fact that such a determination of $\beta_c$ relies
on the absence of non-analiticity points before the strong coupling series of
$M_{\rm d}$ vanishes,
and therefore a non-analiticity at $\beta_c\simeq 0.306$ would invalidate
all strong coupling predictions for $\beta > \beta_c$.

\subsection{Phase distribution of the link operator.}

In 1-d principal chiral models the large-N third order phase
transition is consequence of a compactification of the eigenvalues
of the link operator
\begin{equation}
L \;=\; U(x)U^\dagger (x+\mu)\;,
\end{equation}
which are of the form $\lambda=e^{i\theta}$.
In the weak coupling region ($\beta > \beta_c$) the phase
distribution of the eigenvalues of the link operator $L$,
$\rho(\beta,\theta)$ with $\theta \in (-\pi,\pi]$,
is nonvanishing only in the
region $|\theta|\leq \theta_c(\beta) < \pi$. The third order critical point
$\beta_c$ is determined by the limit condition $\theta_c(\beta)=\pi$,
separating the weak coupling from the strong coupling region where
$\rho (\beta,\pi) > 0$~\cite{Gross}.

In order to see if a similar phenomenon characterizes the
large-$N$ phase transition also in 2-d,
we have extracted from our simulations
the phase distribution  $\rho(\beta,\theta)$
of the eigenvalues of $L$.
Notice that  $\rho(\beta,\theta)=\rho(\beta,-\theta)$ by symmetry,
therefore in the following we will show $\rho(\beta,\theta)$ only in the
range $0\le \theta \le \pi$.
Large-$N$ numerical results seems to support the compactification of the phase
distribution at $\beta_c$, indeed we found
$\rho(\beta,\pi)\simeq 0$
for $\beta \gtrsim \beta_{peak}$ ($\rho(\beta,\pi)$ can be strictly zero only
for $N=\infty$).
This fact is illustrated in
Fig.~\ref{rho21}, where we compare the distributions $\rho(\beta,\theta)$ at
$\beta=0.300$ and $\beta=0.305$ for $N=21$, whose $\beta_{peak}\simeq 0.3025$:
the distribution values at $\theta=\pi$
($\rho(0.300,\pi)\simeq 0.010$ and   $\rho(0.305,\pi)\simeq 0.0007$)
decrease by about a factor 15, becoming very small.
Similar behaviors are observed at the other values of $N$.

In the $SU(N)$ models $\rho(\beta,\theta)$ presents $N$ maxima, as
Fig.~\ref{rho21} shows.
This structure is absent in the $U(N)$ models and should disappear in the
large-$N$ limit,
in that the height of the peaks with respect to the background curve should
vanish.
For example, the $U(N)$ and $SU(N)$ phase distributions at $\beta=0$ are
respectively
\begin{equation}
\rho(0,\theta)\;=\; {1\over 2\pi}\;,
\end{equation}
and
\begin{equation}
\rho(0,\theta) \;=\;
{1\over 2\pi} \left[ 1 + (-1)^{N+1}{2\over N} \cos
\left(N\theta\right)\right]\;.
\label{rhobeta0}
\end{equation}
In our $SU(N)$ simulations we found the peak heights to decrease approximately
as $1/N$.

It is also interesting to see how the distributions $\rho(n,\beta,\theta)$
of the generalized link operators
\begin{equation}
L(n) \;=\; U(x)U^\dagger (x+n\mu)\;,
\end{equation}
($\rho(1,\beta,\theta)\equiv \rho(\beta,\theta)$)
evolve as function of the distance $n$.
In Fig.~\ref{rho15} we plot $\rho(n,\beta,\theta)$
for $N=15$, at $\beta=0.305$ ($\xi_G\simeq 3.79$) and for various values of
$n$.
When $d\equiv n/\xi\rightarrow \infty$, $\rho(n,\beta,\theta)$ appears to tend
to the
$\beta=0$ distribution (\ref{rhobeta0}).

\subsection{Critical slowing down around the large-$N$ singularity.}

The large-$N$ critical behavior causes a phenomenon of critical slowing
down in the Monte Carlo simulations. At sufficiently large $N$ ($N\gtrsim 15$)
and for both $U(N)$ and $SU(N)$ models, the autocorrelation times of the
internal energy
$\tau_E$ and the magnetical susceptibility $\tau_{\chi}$ (estimated by a
blocking procedure)
showed a maximum around the peak of the specific heat, and a sharper and
sharper
behavior with growing $N$.
The increase of the autocorrelation times, with growing $N$, was much
larger around the specific heat peak than elsewhere.
In the $SU(N)$ simulations, $\tau_E$ ($\tau_\chi$) went from
$\sim$ 600 (400) at $\beta=0.3025$ and $N=21$ to
$\sim$ 3000 (2500) at $\beta=0.304$ and $N=30$
(the uncertainty on this numbers is large, they are just indicative).
After the peak of $C$  $\tau_E$ and $\tau_\chi$ decreased,
for example at $N=30$ and $\beta=0.305$ $\tau_E\simeq 700$ and $\tau_\chi\simeq
300$.
Similar behavior was observed in the $U(N)$ simulations.
The above critical slowing down phenomenon represents the most serious
difficulty in getting
numerical results around $\beta_c$ at larger $N$
by the Monte Carlo algorithm used in this work.
At large correlation length $\tau_\chi$ increases again
due the critical slowing down associated to the continuum limit,
while $\tau_E$ tends to be stable.

We want to mention an attempt for a better algorithm in the $U(N)$ case, by
constructing a microcanonical updating
involving  globally the $U(N)$ matrix instead of using its subgroups.
A  microcanonical updating of $U$ according to the action
\begin{equation}
A(U)\;=\; {\rm Re} \,{\rm Tr} \left[U F\right]
\end{equation}
can be achieved by performing the reflection with respect to the
$U(N)$ matrix $U_{max}$ which maximizes $A(U)$:
\begin{eqnarray}
U_{new}&=& U_{max}\, U_{old}^\dagger \,U_{max}\;,\nonumber \\
U_{max}&=& {1\over \sqrt{F^\dagger F} } F^\dagger\;.
\label{mcup}
\end{eqnarray}
Notice that the determination of $U_{max}$
requires the diagonalization of the complex matrix $F$.
The update~(\ref{mcup}) does not change the action and it must be combined
with ergodic algorithms (e.g. heat bath).
We found that, at large $N$ and in the region of $\beta$ values we considered,
the algorithm based on the $SU(2)$ and $U(1)$ subgroups
performs better than those based on the updating (\ref{mcup}).
The latter  may become convenient at relatively small $N$ and/or for larger
correlation lengths.
On the other hand, at large space correlation lengths
multigrid algorithms should eventually become more efficient, in that they
should
have smaller dynamical exponents
(see Refs.~\cite{Sokal,Meyer} for some implementations of
multigrid algorithms in the context of lattice chiral  models).

\appendix{}
\label{AA}

The free energy of 1-d $U(N)$ lattice chiral models
can be written in terms of a determinant of modified Bessel functions
\begin{equation}
F(N,\beta)\;=\; {1\over N^2} \ln Z(N,\beta)\;=\;
{1\over N^2} \ln {\rm det} \,I_{j-i}(2N\beta)\;.
\label{freee}
\end{equation}
The specific heat can be obtained by
\begin{equation}
C(N,\beta)\;=\; {1\over N}{{\rm d} E\over {\rm d} T}\;=\;
{1\over 2} \beta^2 {{\rm d}^2 F\over {\rm d}\beta^2}\;.
\label{cv}
\end{equation}
The large-$N$ limit of the specific heat shows the existence of
a third order phase transition at $\beta_c=1/2$,
indeed we have
\begin{eqnarray}
C(\infty,\beta)&=& \beta^2 \;\;\;\;\;\;\;\; {\rm for}\;\;\;\beta\leq
\beta_c\;,\nonumber \\
C(\infty,\beta)&=& {1\over 4} \;\;\;\;\;\;\;\; {\rm for}\;\;\;\beta\geq
\beta_c.
\label{Cvinf}
\end{eqnarray}
The singularity at $\beta_c$ can be characterized by a critical exponent
$\alpha=-1$. From the double scaling limit of the corresponding
unitary matrix model the correlation length exponent turns out to be
$\nu=3/2$~\cite{Periwal}.
$\alpha$ and $\nu$ satisfy a
hyperscaling relationship associated to a two-dimensional
critical phenomenon: $2\nu=2-\alpha$.

1-d $U(N)$ lattice chiral models present a peak in the specific heat,
whose position $\beta_{peak}(N)$ should approach $\beta_c$ with increasing $N$.
The finite $N$ scaling arguments already mentioned in this paper lead to the
Ansatz (\ref{FNS})  for the positions of the specific heat peaks.
In Table~\ref{C1dim} we report the values of $\beta_{peak}(N)$ and
$C(N,\beta_{peak})$
as function of $N$ up to $N=11$.
As shown in Fig.~\ref{bpeak1dim}, the large N behavior of $\beta_{peak}(N)$
is well fitted by
\begin{equation}
\beta_{peak}(N)\;=\;\beta_c\,+\,aN^{-\epsilon} \,+\,bN^{-2\epsilon}
\label{bpkdim1}
\end{equation}
with $\epsilon=2/3$, and therefore $\nu=1/\epsilon=3/2$
($a\simeq 0.595$ and $b\simeq 0.13$).
The result $\nu=3/2$ was also found in the finite $N$ scaling of the partition
function zeroes~\cite{Heller}.



\figure{
Specific heat vs.\ $\beta$ for $SU(N)$ models.
The solid line represents
the strong-coupling determination, whose estimate of the
critical $\beta$ is indicated by the vertical dashed lines.
The thick solid lines above the peaks represent our estimates of
$\beta_{peak}$.
\label{CUN}}

\figure{
Specific heat vs.\ $\beta$ for $U(N)$ models.
The solid line represents
the strong-coupling determination, whose estimate of the
critical $\beta$ is indicated by the vertical dashed lines.
The large solid lines above the peaks
represent our estimates of $\beta_{peak}$.
\label{CSUN}}

\figure{
$\beta_{peak}(N)$ vs.\ $1/N$.
The dashed lines show the fit result.
\label{bpeak}}

\figure{
$\xi_G$ vs.\ $\beta$.
The solid line represents
the strong-coupling determination, whose estimate of the
critical $\beta$ is indicated by the vertical dashed lines.
\label{xi}}

\figure{
$\chi_{\rm d}$ vs.\ $\beta$.
The vertical dashed line indicates the estimate of $\beta_c$.
The filled symbols indicate the positions of the peak
of $C$ at $N=9,15,21$.
\label{chid}}

\figure{
$\rho(\beta,\theta)$ for the $SU(21)$ model at $\beta=0.300$
and $\beta=0.305$.
\label{rho21}}

\figure{
$\rho(n,\beta,\theta)$ for the $SU(15)$ model at $\beta=0.305$
for various value of $n$.
\label{rho15}}

\figure{
$\beta_{peak}(N)$ vs.\ $1/N$ in 1-d $U(N)$ lattice chiral models.
\label{bpeak1dim}}


\begin{table}
\squeezetable
\caption{Numerical results for $U(N)$. }
\label{UNtable}
\begin{tabular}{cr@{}lccr@{}lr@{}lr@{}lr@{}lr@{}lr@{}lr@{}l}
\multicolumn{1}{c}{$N$}&
\multicolumn{2}{c}{$\beta$}&
\multicolumn{1}{c}{$L$}&
\multicolumn{1}{c}{${\rm Stat}$}&
\multicolumn{2}{c}{$E$}&
\multicolumn{2}{c}{$C$}&
\multicolumn{2}{c}{$\chi$}&
\multicolumn{2}{c}{$\xi_G$}&
\multicolumn{2}{c}{$\chi_{\rm d}$}&
\multicolumn{2}{c}{$\xi_{{\rm d}}$}\\
\tableline
9 & 0&.30   & 24& 100k & 0&.60374(8) & 0&.284(6)    & 10&.50(3)  & 2&.058(14)&
1&.64(2) & 0&.6(2)\\
  & 0&.31   & 30& 150k & 0&.56706(9) & 0&.427(11)   & 15&.43(4)  & 2&.649(17)&
2&.82(4) & 1&.0(2)\\
  & 0&.313  & 30& 300k & 0&.55215(10)& 0&.541(11)   & 18&.53(5)  & 2&.972(14)&
3&.93(5) & 1&.2(2)\\
  & 0&.315  & 36& 150k & 0&.54077(14)& 0&.61(2)     & 21&.80(10) & 3&.36(3)  &
5&.59(9) & 1&.9(2)\\
  & 0&.318  & 36& 300k & 0&.52128(10)& 0&.66(2)     & 29&.47(9)  & 4&.08(3)
&11&.3(2)  & 3&.0(2)\\
  & 0&.3185 & 36& 400k & 0&.51799(11)& 0&.69(2)     & 31&.17(11) & 4&.22(2)
&12&.8(2)  & 3&.2(2)\\
  & 0&.319  & 42& 300k & 0&.51482(11)& 0&.69(2)     & 32&.86(13) & 4&.38(3)
&14&.7(4)  & 3&.3(3)\\
  & 0&.320  & 42& 400k & 0&.50816(10)& 0&.66(2)     & 37&.2(2)   & 4&.73(3)
&20&.5(5)  & 4&.3(3)\\
  & 0&.323  & 48& 330k & 0&.49172(9) & 0&.50(2)     & 51&.6(3)   & 5&.83(4)
&55&(3)    & 8&.5(5)\\
  & 0&.323  & 60& 200k & 0&.49166(11)& 0&.52(2)     & 51&.7(3)   & 5&.79(7)
&57&(3)    & 7&.7(7)\\\hline

15& 0&.28   & 18& 150k & 0&.65373(4)  & 0&.163(3)  &  6&.924(7) & 1&.519(4) &
1&.030(8) & &\\
  & 0&.30   & 24& 200k & 0&.60276(6)  & 0&.300(10) & 10&.81(2)  & 2&.063(9) &
1&.29(2)  & &\\
  & 0&.305  & 24& 180k & 0&.58405(9)  & 0&.396(12) & 13&.09(3)  & 2&.346(9) &
1&.57(2)  & &\\
  & 0&.308  & 30& 250k & 0&.56875(9)  & 0&.57(2)   & 15&.55(3)  & 2&.632(10)&
2&.14(2)  & 0&.7(2)\\
  & 0&.310  & 30& 300k & 0&.55423(12) & 0&.78(3)   & 18&.83(5)  & 2&.996(11)&
3&.31(5)  & 1&.2(2) \\
  & 0&.311  & 30& 500k & 0&.54453(10) & 0&.97(4)   & 21&.58(4)  & 3&.276(8) &
4&.87(6)  & 1&.6(1)\\
  & 0&.311  & 36& 300k & 0&.54470(13) & 0&.97(4)   & 21&.52(6)  & 3&.276(14)&
4&.85(9)  & 1&.5(2)\\
  & 0&.312  & 36& 600k & 0&.53374(11) & 1&.05(4)   & 25&.57(10) & 3&.67(2)  &
8&.42(15) & 2&.8(2)\\
  & 0&.313  & 36& 500k & 0&.52365(12) & 0&.94(4)   & 30&.42(10) &
4&.131(15)&14&.5(5)   & 3&.7(3)\\
  & 0&.315  & 42& 200k & 0&.50920(10) & 0&.50(3)   & 40&.0(2)   & 4&.95(4)
&42&(5)     & 7&.2(7)\\
  & 0&.315  & 48& 300k & 0&.50915(7)  & 0&.49(2)   & 39&.7(2)   & 4&.89(4)
&46&(3)     & 8&.1(7)\\\hline

21& 0&.28   & 18& 100k & 0&.65373(4) & 0&.162(4)  &  6&.972(8) & 1&.526(5)  &
0&.991(8) & &\\
  & 0&.30   & 24& 200k & 0&.60273(6) & 0&.303(9)  & 10&.881(13)& 2&.069(6)  &
1&.140(8) & &\\
  & 0&.3025 & 24& 300k & 0&.59390(6) & 0&.361(10) & 11&.869(14)& 2&.185(6)  &
1&.220(9) & &\\
  & 0&.305  & 30& 300k & 0&.58318(6) & 0&.446(14) & 13&.31(2)  & 2&.364(9)  &
1&.394(11)& & \\
  & 0&.308  & 30& 200k & 0&.56337(15)& 0&.88(6)   & 16&.79(5)  & 2&.748(13) &
2&.15(3)  & & \\
  & 0&.309  & 30& 300k & 0&.5512(3)  & 1&.31(13)  & 19&.92(10) & 3&.09(2)   &
3&.60(9)  & 1&.2(2)\\
  & 0&.3095 & 30& 450k & 0&.5415(3)  & 1&.98(15)  & 23&.11(13) & 3&.43(2)   &
6&.6(3)   & 2&.4(2)\\
  & 0&.31   & 36& 300k & 0&.5337(2)  & 1&.28(10)  & 26&.23(11) & 3&.75(2)
&11&.1(5)   & 3&.4(3)\\
\end{tabular}
\end{table}

\begin{table}
\squeezetable
\caption{Numerical results for $SU(N)$.
When more than one lattice size appears,
the corresponding results were obtained collecting
data of simulations at the reported lattice sizes
(which were, in all cases, in agreement within the errors).}
\label{SUNtable}
\begin{tabular}{cr@{}lccr@{}lr@{}lr@{}lr@{}lr@{}l}
\multicolumn{1}{c}{$N$}&
\multicolumn{2}{c}{$\beta$}&
\multicolumn{1}{c}{$L$}&
\multicolumn{1}{c}{${\rm Stat}$}&
\multicolumn{2}{c}{$E$}&
\multicolumn{2}{c}{$C$}&
\multicolumn{2}{c}{$\chi$}&
\multicolumn{2}{c}{$\xi_G$}\\
\tableline
9 &  0&.290 & 30 & 200k & 0&.58774(8)   & 0&.412(7)    &13&.32(3)    &
2&.353(11)\\
  &  0&.294 & 30,36 & 600k & 0&.56788(6)  & 0&.435(6)   &16&.89(3)    &
2&.793(14)\\
  &  0&.295 & 24,30,36,42 & 900k & 0&.56284(4)   & 0&.443(5)    &18&.00(2)    &
2&.910(9)  \\
  &  0&.2955& 36 & 500k & 0&.56026(5)   & 0&.442(6)    &18&.58(4)    & 2&.95(2)
 \\
  &  0&.296 & 30,36 & 600k & 0&.55781(5)   & 0&.438(6)    &19&.20(3)    &
3&.03(2) \\
  &  0&.2965& 36 & 600k & 0&.55531(6)   & 0&.436(6)    &19&.86(4)    & 3&.08(2)
\\
  &  0&.300 & 30,36,42 & 350k & 0&.53846(9)  & 0&.413(11)   &25&.27(7)   &
3&.66(2)  \\
  &  0&.310 & 42,48,54 & 500k & 0&.50030(4)  & 0&.306(5)    &47&.25(12)  &
5&.43(3)  \\\hline

15&  0&.295 & 24 & 200k & 0&.60013(11) & 0&.47(2)      &11&.47(2)    &
2&.149(9) \\
  &  0&.299 & 30 & 300k & 0&.57564(10) & 0&.66(2)      &15&.07(3)    &
2&.577(11)  \\
  &  0&.300 & 24 & 400k & 0&.56798(10) & 0&.69(4)      &16&.55(3)    &
2&.738(6)  \\
  &  0&.300 & 30 & 400k & 0&.56805(10) & 0&.70(3)      &16&.57(3)    &
2&.746(9)  \\
  &  0&.300 & 36 & 600k & 0&.56807(9)  & 0&.66(2)      &16&.58(2)    &
2&.745(8)  \\
  &  0&.300 & 42 & 500k & 0&.56810(5)  & 0&.70(2)      &16&.57(3)    &
2&.752(12)  \\
  &  0&.3005& 36 & 600k & 0&.56430(7)  & 0&.68(2)      &17&.41(3)    &
2&.833(11)  \\
  &  0&.301 & 36 & 500k & 0&.56054(6)  & 0&.68(2)      &18&.31(3)    &
2&.940(10) \\
  &  0&.302 & 36 & 500k & 0&.55300(5)  & 0&.65(2)      &20&.26(3)    &
3&.131(9) \\
  &  0&.305 & 36 & 500k & 0&.53418(6)  & 0&.516(13)    &26&.86(6)    &
3&.786(11) \\
  &  0&.310 & 45 & 300k & 0&.51178(4)  & 0&.354(7)     &39&.06(10)   & 4&.80(2)
\\\hline

21&  0&.300 & 24 & 300k & 0&.58810(10) & 0&.65(3)      &12&.90(2)   & 2&.310(6)
\\
  &  0&.302 & 30 & 500k & 0&.57049(13) & 1&.00(4)      &15&.91(3)   & 2&.665(7)
\\
  &  0&.302 & 36 & 600k & 0&.57069(8)  & 0&.95(3)      &15&.87(2)   & 2&.659(8)
\\
  &  0&.3025& 24 & 400k & 0&.56490(20) & 1&.14(6)      &17&.09(6)   & 2&.787(8)
\\
  &  0&.3025& 30 & 400k & 0&.56517(14) & 1&.02(5)      &17&.02(4)   & 2&.784(8)
\\
  &  0&.3025& 36 & 500k & 0&.56491(11) & 1&.04(5)      &17&.11(4)   &
2&.800(10) \\
  &  0&.303 & 36 & 500k & 0&.55959(9)  & 0&.96(4)      &18&.38(3)   & 2&.936(9)
\\
  &  0&.305 & 30 & 500k & 0&.54100(8)  & 0&.72(2)      &24&.14(5)   & 3&.526(8)
\\
  &  0&.310 & 42 & 240k & 0&.51548(6)  & 0&.41(2)      &36&.66(12)  & 4&.61(2)
\\\hline

30&  0&.300 & 24 & 150k & 0&.59927(8)  & 0&.38(2)      &11&.35(2)   & 2&.114(7)
\\
  &  0&.3025& 30 & 200k & 0&.58479(10) & 0&.79(5)      &13&.24(3)   & 2&.338(7)
\\
  &  0&.303 & 30 & 200k & 0&.58007(15) & 1&.00(8)      &13&.99(4)   &
2&.433(10) \\
  &  0&.304 & 24 & 500k & 0&.5625(4)   & 2&.4(3)       &17&.55(7)   &
2&.857(10) \\
  &  0&.304 & 30 & 500k & 0&.5632(3)   & 2&.3(2)       &17&.40(7)   & 2&.829(8)
\\
  &  0&.305 & 30 & 200k & 0&.5466(2)   & 1&.05(10)     &22&.13(5)   &
3&.320(12) \\
\end{tabular}
\end{table}

\begin{table}
\caption{$\beta_{peak}(N)$ and $C(N,\beta_{peak})$ versus $N$
for 1-d $U(N)$ lattice chiral models.}
\label{C1dim}
\begin{tabular}{cr@{}lr@{}l}
\multicolumn{1}{c}{$N$}&
\multicolumn{2}{c}{$\beta_{peak}$}&
\multicolumn{2}{c}{$C(\beta_{peak})$}\\
\tableline
2 &  0.&930889 & 0.&29461215 \\
3 &  0.&818356 & 0.&27992604 \\
4 &  0.&758001 & 0.&27269388 \\
5 &  0.&719664 & 0.&26839003 \\
6 &  0.&692846 & 0.&26553250 \\
7 &  0.&672876 & 0.&26349442 \\
8 &  0.&657337 & 0.&26196545 \\
9 &  0.&644848 & 0.&26077452 \\
10&  0.&634554 & 0.&25981956 \\
11&  0.&625899 & 0.&25903594 \\
\end{tabular}
\end{table}

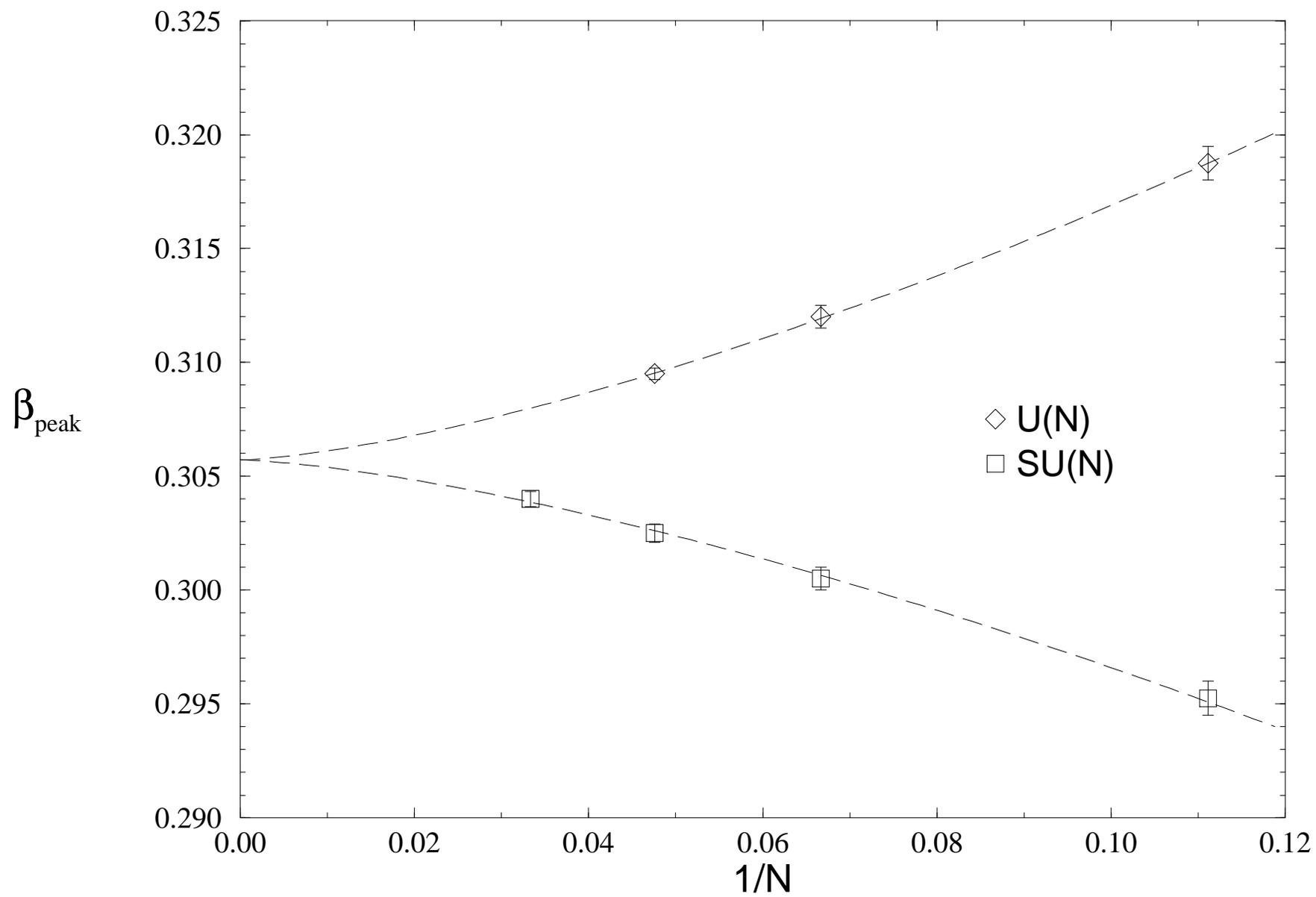
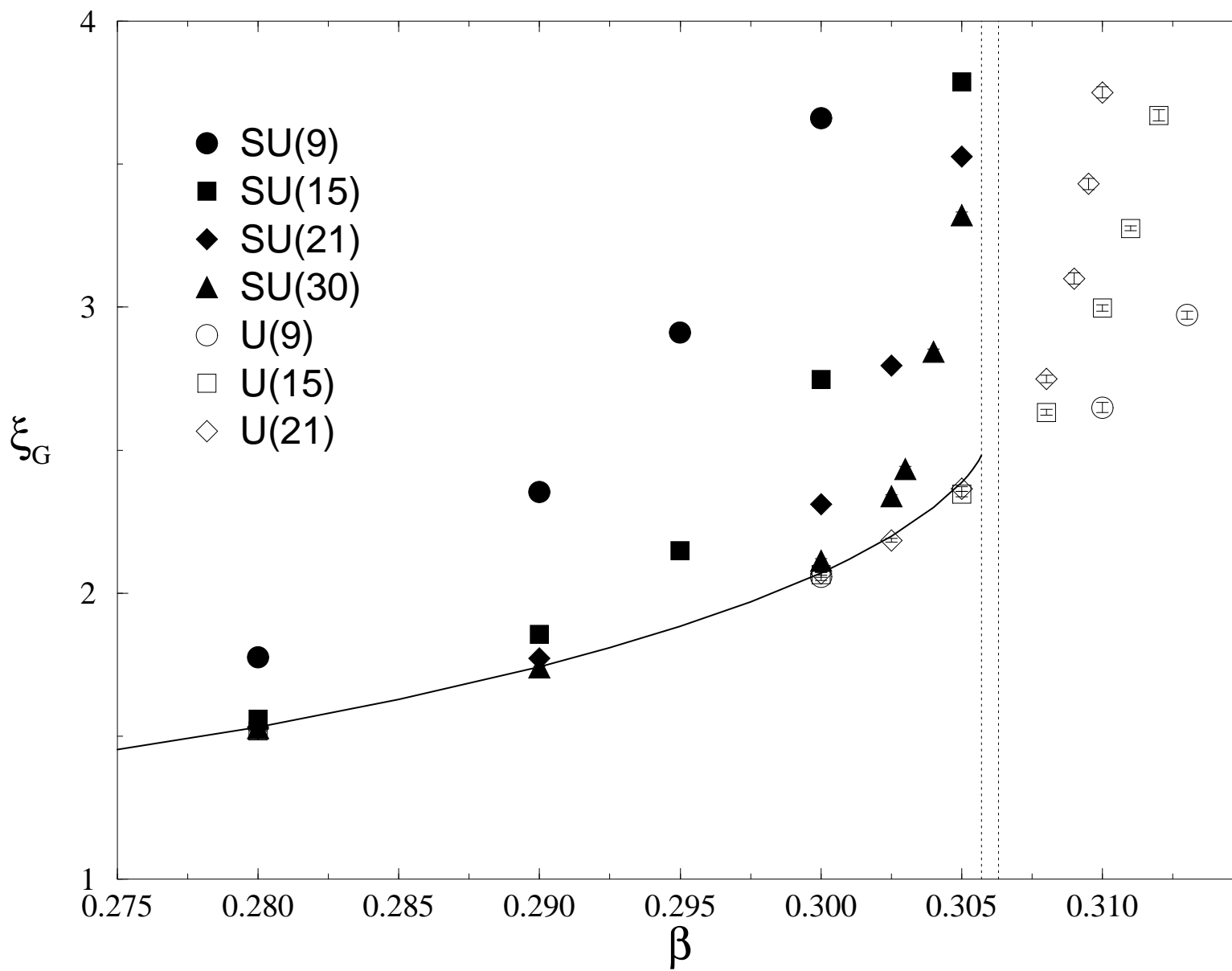
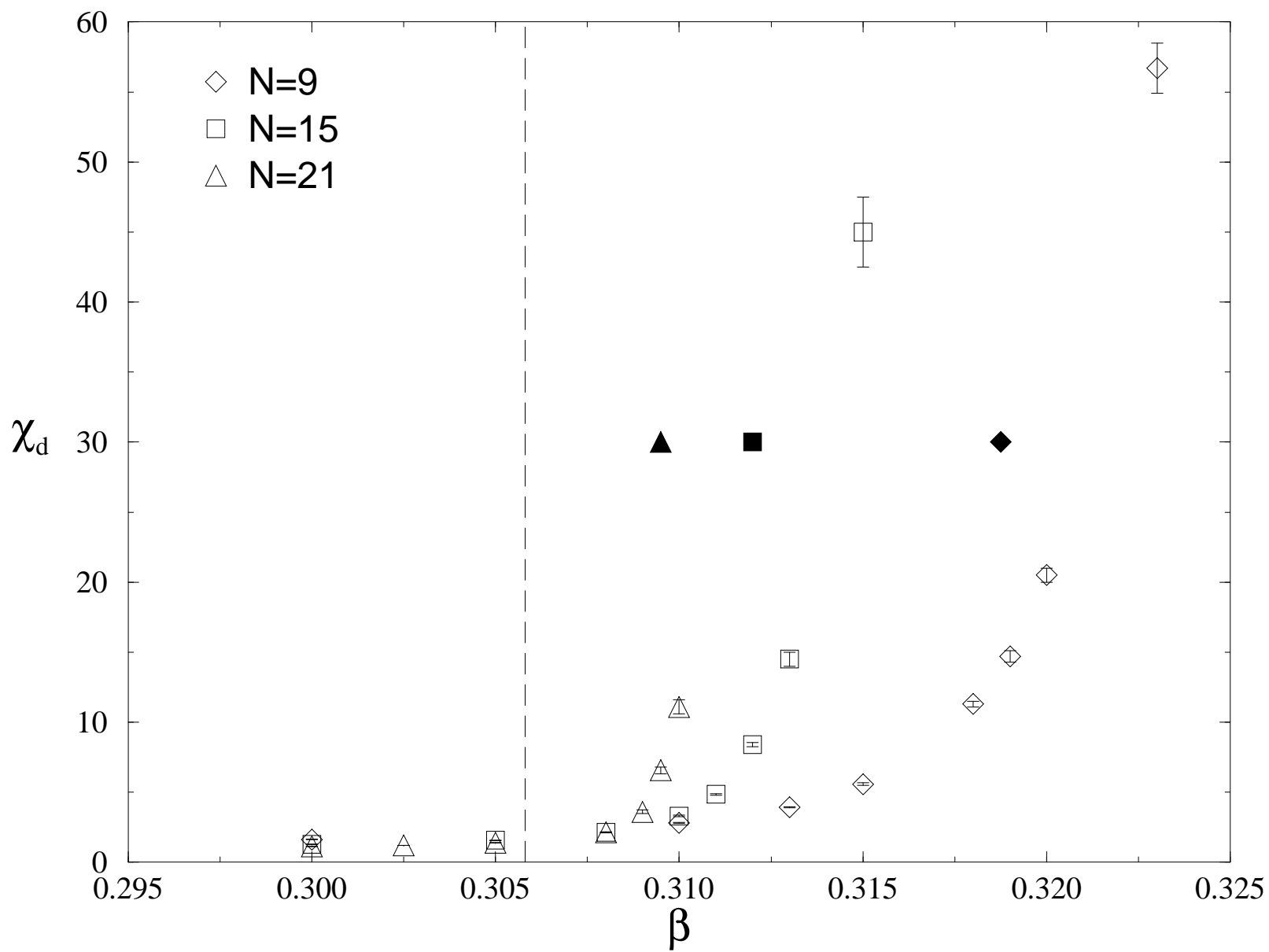
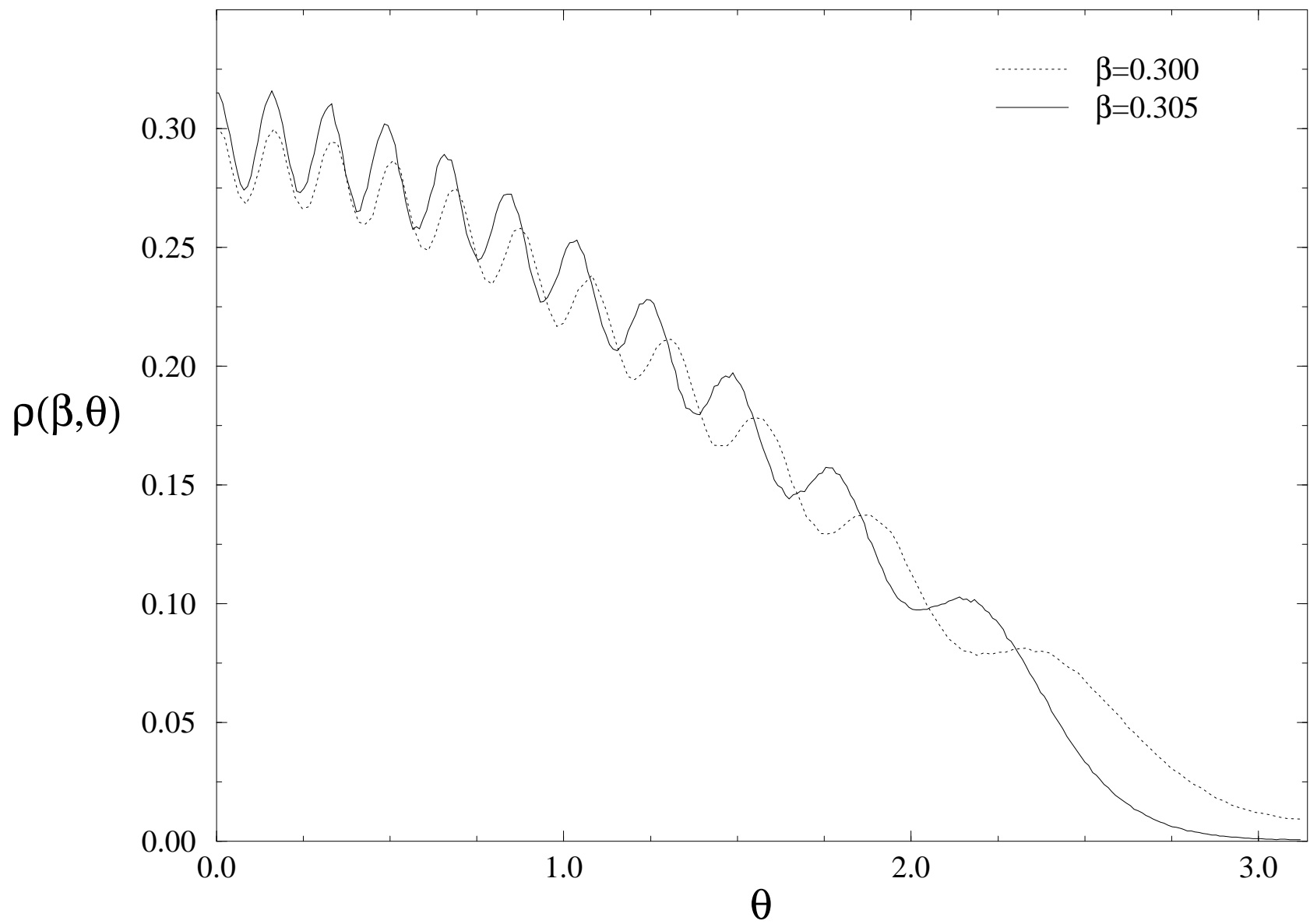
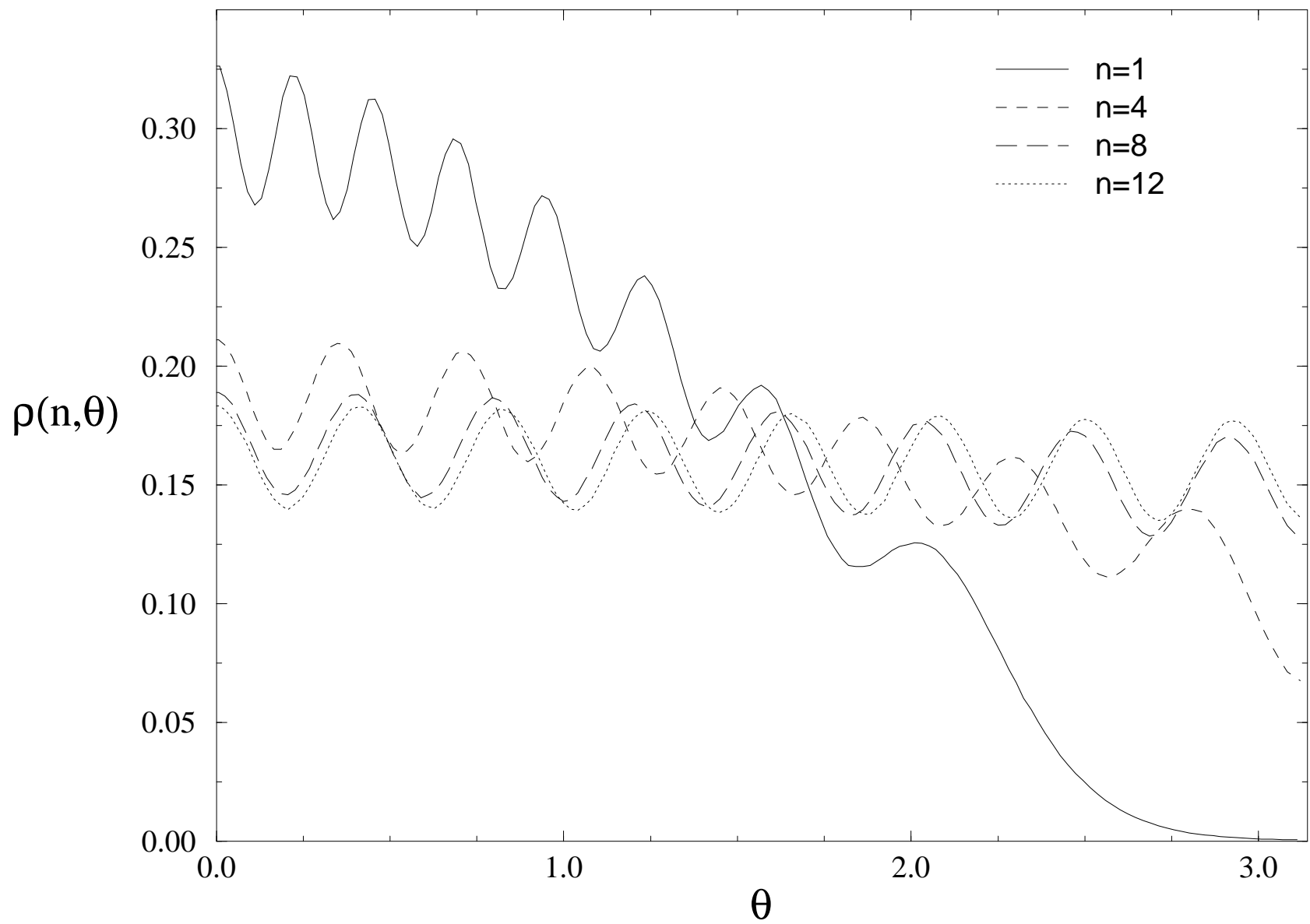
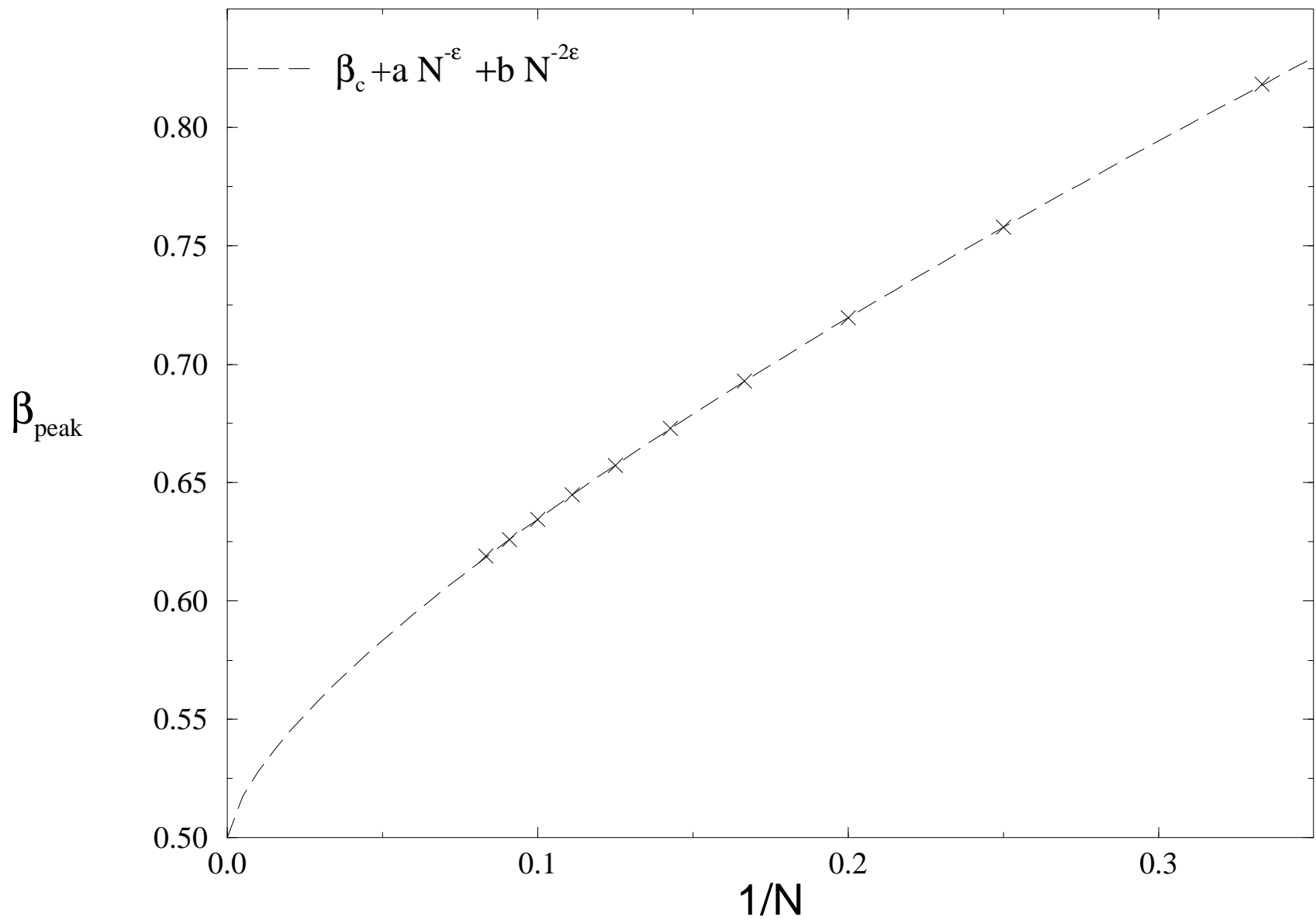
\end{document}